\documentclass[9pt,twocolumn,twoside]{osajnl}
\usepackage{setspace}
\usepackage[utf8]{inputenc}
\usepackage[T1]{fontenc}
\usepackage{cleveref}
\usepackage[detect-weight]{siunitx}
\usepackage{relsize}
\usepackage[nomain, acronym]{glossaries}
\usepackage{dblfloatfix}
\setacronymstyle{long-sm-short}
\glsdisablehyper

\makeatletter
\newcommand\etc{etc\@ifnextchar.{}{.\@}}
\newcommand\etal{\emph{et al}\@ifnextchar.{}{.\@}}
\makeatother

\newacronym{CMOS}{CMOS}{complementary metal-oxide-semiconductor}
\newacronym{SDM}{SDM}{spatial division multiplexing}
\newacronym{SOI}{SOI}{silicon-on-insulator}
\newacronym{SOITEC}{SOITEC}{SOITEC}\glsunset{SOITEC}
\newacronym{SR}{SR}{split ratio}
\newacronym{IL}{IL}{insertion loss}
\newacronym{XT}{XT}{crosstalk}
\newacronym{SPINS}{SPINS}{Inverse Design Software for Nanophotonic Structures}\glsunset{SPINS}
\newacronym{SEM}{SEM}{scanning electron microscopy}
\newacronym{OSA}{OSA}{optical spectrum analyzer}
\newacronym{FDTD}{FDTD}{finite difference time domain}
\newacronym{PECVD}{PECVD}{plasma-enhanced chemical vapor deposition}
\newacronym{RIE}{RIE}{reactive ion etching}
\newacronym{ANT}{ANT}{Applied Nanotools}
\newacronym{MIMO}{MIMO}{multiple-input multiple-output}
\journal{ol} 

\setboolean{shortarticle}{true}

\title{Compact Dual-Polarization Silicon Integrated Couplers for Multicore Fibers}

\author[1,2,3,*]{Julian L. Pita Ruiz}
\author[1,2]{Lucas G. Rocha}
\author[4]{Jun Yang}
\author[4]{\c{S}\"ukr\"u Ekin Kocaba\c{s}}
\author[4]{Ming-Jun Li}
\author[1,5]{Ivan Aldaya}
\author[1,3,4]{Paulo Dainese}
\author[1,2]{Lucas H. Gabrielli}
\affil[1]{Photonics Research Center, University of Campinas, Campinas 13083-859, SP, Brazil}
\affil[2]{School of Electrical and Computer Engineering, University of Campinas, Campinas 13083-852, SP, Brazil}
\affil[3]{``Gleb Wataghin'' Physics Institute, University of Campinas, Campinas 13083-859, SP, Brazil}
\affil[4]{Corning Research \& Development Corporation, One Science Drive, Corning, New York 14830, USA}
\affil[5]{Campus of São João da Boa Vista, State University of São Paulo, São João da Boa Vista 13876-750, SP, Brazil}
\affil[*]{Corresponding author: pita@ifi.unicamp.br}

\ociscodes{
(050.1950) Diffraction gratings; (060.2330) Fiber optics communications; (130.3120) Integrated optics devices.}

\begin{abstract}
Compact fiber-to-chip couplers play an important role in optical interconnections, especially in data centers.
However, the development of couplers has been mostly limited to standard single-mode fibers, with few devices compatible with multicore and multimode fibers.
Through the use of state-of-the-art optimization algorithms, we designed a compact dual-polarization coupler to interface chips and dense multicore fibers, demonstrating, for the first time, coupling to both polarizations of all the cores, with measured coupling efficiency of \SI{-4.3}{dB} and with a \SI{3}{dB} bandwidth of \SI{48}{nm}.
The dual-polarization coupler has footprint of $\SI{200}{\um^2}$ per core, which makes it the smallest fiber-to-chip coupler experimentally demonstrated on a standard silicon-on-insulator platform.
\end{abstract}

\setboolean{displaycopyright}{true}

\begin{document}

\maketitle

\Gls{SDM} based on multicore fibers has emerged as one of the potential candidates to overcome the capacity crunch of traditional optical communication systems~\cite{zhang2020enabling, mendinueta2020high}.
Nonetheless, the adoption of this kind of fiber will be possible only if light can be efficiently coupled to all the cores.
So far, the widely adopted solution relies on fiber fan-outs that connect several single-core fibers to a single  multicore fiber~\cite{downie2020examining, gan2019ultra, shikama2018multicore}.
When connecting to an integrated photonics transceiver, however, grating couplers offer a more straightforward and compact alternative, directly coupling the multicore fiber to the integrated waveguides.
Focused grating couplers can efficiently couple a single polarization from a single-mode fiber into the chip~\cite{mekis2010grating, marchetti2017high}, and can also be applied to multicore fibers with large core-to-core separation if only one polarization is employed.
Losing polarization diversity, however, is not acceptable for high-end communication systems, and therefore it is paramount to address both polarization states of each core.
This open problem has motivated an increasing effort to the development of couplers with ultra-compact footprints, without impacts on coupling efficiency, wavelength bandwidth, and \gls{XT}.

\begin{figure*}[htb]
\centering
\includegraphics[scale=1.2]{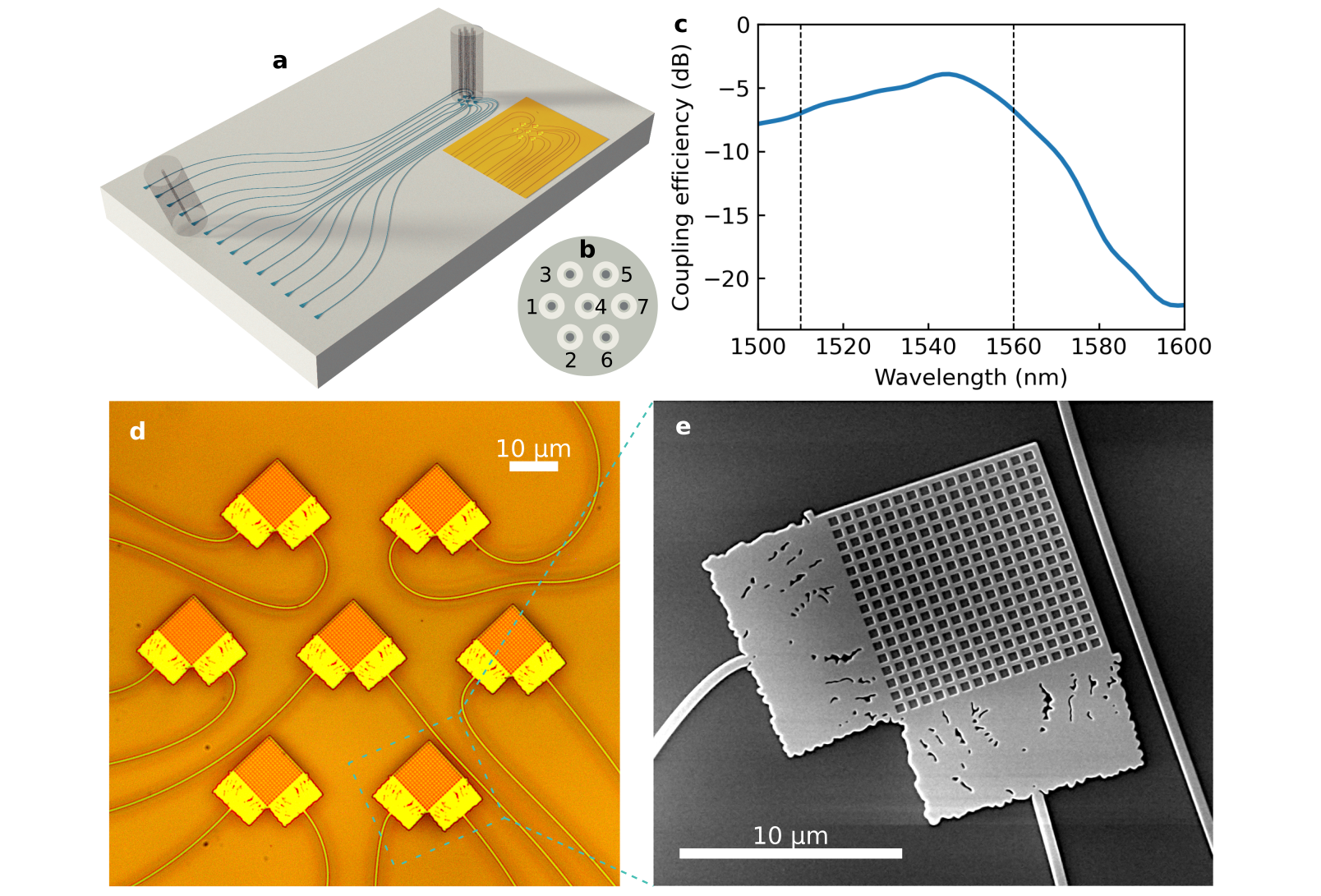}
\caption{(a)~Illustration of the integrated fan-out to test the compact dual-polarization coupler.
(b)~Image of the cross-section of the seven-core fiber.
(c)~Simulated coupling efficiency of the dual-polarization coupler.
The dashed lines indicate the \num{3}-\si{dB} bandwidth of the device.
(d)~Optical microscopy image of the seven-core dual-polarization coupler fabricated on SOI technology.
(e)~\Gls{SEM} image of one of the manufactured 2d grating alongside its compact tapers.}
\label{fig:2d}
\end{figure*}

Significant progress has been made recently in couplers for fibers with large inter-core separation.
For example, Dwivedi~\etal{} reported a single-polarization compact focused grating employing silicon nitride instead of silicon.
The lower index contrast between the silicon-nitride core and silica cladding resulted in weaker modal confinement, thus reducing the footprint of the grating down to $\SI{16}{\um}\times\SI{18}{\um}$, achieving a \num{3}-\si{dB} bandwidth of \SI{60}{nm}, a \gls{XT} of \SI{-35}{dB}, and a coupling efficiency of \SI{-5.4}{dB}~\cite{dwivedi_integrated_2017,dwivedi_demonstration_2017,dwivedi_multicore_2018}.
Ding~\etal{} demonstrated another single-polarization coupler with \SI{-3.8}{dB} efficiency, \SI{48}{nm} bandwidth, and \SI{-32}{dB} \gls{XT}~\cite{ding_-chip_2015}.
In this case, the improved efficiency is achieved by placing a metal layer below the grating, which is not be desirable due to fabrication complexity and incompatibility with conventional foundry processes~\cite{ding_-chip_2015,ding_experimental_2016, ding_high-dimensional_2017, ding_reconfigurable_2016}.
Meanwhile, Tong~\etal{} proposed a single-polarization coupler with an efficiency of \SI{-2.8}{dB} and a bandwidth of \SI{58.6}{nm} through the use of an extra poly-silicon layer above the silicon layer~\cite{tong_efficient_nodate, tong_112-tbits_2020}, requiring, however, a relatively large inter-core spacing of \SI{45}{\um}.
No measurement of \gls{XT} is available for that coupler, because the experimental characterization was performed using a single-core fiber.
Regarding dual-polarization coupling to fibers with more than four cores, Hayashi~\etal{}~\cite{hayashi_end--end_2017} reported coupling to half of the cores of a $2\times4$ multicore fiber with \SI{45}{\um} core separation, limited primarily by the coupler footprint.
Because the footprint limits the integration density of the individual couplers, the other half of the fiber cores can only be coupled through single polarization structures.
To the best of our knowledge, no dual-polarization coupler proposal has so far been able to address all core of a multicore fiber, which is required for transmission at full capacity.

In this work, we present a compact dual-polarization silicon coupler capable of addressing all seven-cores on a hexagonal arrangement with center-to-center distance of \SI{32}{\um}, illustrated in \cref{fig:2d}a.
The device shows coupling efficiency of \SI{-4.3}{dB} with a \SI{3}{dB} bandwidth of \SI{48}{nm} and \gls{XT} below \SI{-42.7}{dB} between adjacent cores.

The coupler is composed of seven two-dimensional gratings with area $\SI{10}{\um}\times\SI{10}{\um}$ connected via ultra-compact tapers to single-mode waveguides.
\Cref{fig:2d}e shows a \gls{SEM} image of one of the gratings, each of which is formed by the superposition of two identical single-polarization gratings~\cite{lacava2016design,watanabe20192}, designed and simulated by the \gls{FDTD} method. 
Different devices were fabricated and characterized, revealing that the best found performance corresponded to a pitch of \SI{623}{nm} and a fill factor of 0.7. 
The tapers connecting each grating output to a \SI{450}{nm}-width silicon waveguide were designed using \gls{SPINS}~\cite{Su_2020_Nanophotonic} on an area of only $\SI{10}{\um}\times\SI{5}{\um}$.
Minimal feature size was set to \SI{100}{nm} in compliance with the fabrication process design kit to ensure high fabrication yield~\cite{piggott2017fabrication}.
Simulations indicate an insertion loss of \SI{0.5}{dB} for the taper.
Such a low insertion loss is an excellent result considering the compact footprint is key to allow the network of 14 single-mode waveguides to be routed around all fiber core positions without introducing bend losses, as \cref{fig:2d}d shows.
Because the gratings are designed for a \SI{10}{\degree} coupling angle, the 2-dimensional gratings and tapers are not exactly orthogonal, but at an angle of \SI{83.1}{\degree}, which was optimized through 3D simulations. 
The complete testing system for the dual-polarization coupler is illustrated in \cref{fig:2d}a, where each waveguide is coupled to a single-mode fiber through a conventional focusing grating.
The experimental characterization was performed in fan-out configuration, that is, the light is injected using single-core fibers and collected at the multi-core fiber. It is worth mentioning that the proposed coupler can be equally employed in fan-in configuration. In this case, depending on the particular application, the light coupled to the waveguides corresponding to the two polarizations can be combined in the optical domain or individually detected for \gls{MIMO} processing.
The 3D \gls{FDTD} simulated coupling efficiency for each mode is \SI{-3.9}{dB} with \SI{3}{dB} bandwidth of \SI{50}{nm}, as can be seen in the plot of \cref{fig:2d}c.

The device was fabricated in a dedicated run at \gls{ANT} facilities using a \gls{SOI} wafer from \gls{SOITEC} with a buried oxide layer of \SI{3}{\um}, and a silicon layer of \SI{250}{nm}.
Patterning was done via electron-beam lithography with minimal feature sizes for both positive and negative layers of \SI{100}{nm}.
Two lithography steps (\SI{250}{nm} full etch and \SI{120}{nm} partial etch) in conjunction with reactive ion etching was used to transfer patterns to the silicon layer.
As a final step, \SI{1}{\um} of silicon dioxide cladding was deposited on the samples by \gls{PECVD}.

The multicore fiber used has \SI{200}{m} in length and seven identical cores (shown in \cref{fig:2d}b) with modal diameters of \SI{8}{\um} at \SI{1550}{nm}.
The center-to-center separation between adjacent cores is \SI{32}{\um} and their \gls{XT}, measured over a length of \SI{10}{km}, is of \SI{-43}{dB}.
This fiber was aligned with the dual-polarization coupler on one side of the chip, while an array of 14 focused grating couplers was used on the other side to allow coupling to single-mode fibers.
The best coupling efficiency was achieved when the multi-core fiber had an angle of approximately \SI{13}{\degree} with respect to the vertical.
Light from a tunable continuous laser source was coupled to the single-mode fiber and collected at the end of the multicore fiber.
For coupling efficiency measurements, the total power at the output of the multicore fiber was measured employing a power meter equipped with an integrating sphere, whereas to measure the \gls{XT}, a second single-mode fiber was used.
This second fiber was aligned to each one of the adjacent cores with the help of two piezo-controlled stages and delivered the power to an \gls{OSA}.
The efficiency of the focused grating couplers for single-core fibers and the waveguide propagation losses were measured through linear regression of 8 calibration waveguides (2 waveguides of \SI{0.5}{cm}, 2 waveguides of \SI{1}{cm}, 2 waveguides of \SI{3}{cm}, and 2 waveguides of \SI{5}{cm}).
The obtained values were \SI{-3.65}{dB} for the coupling efficiency and \SI{1.4}{\dB/cm} for the propagation loss.

\begin{figure}[htb]
\centering
\includegraphics[scale=1.2]{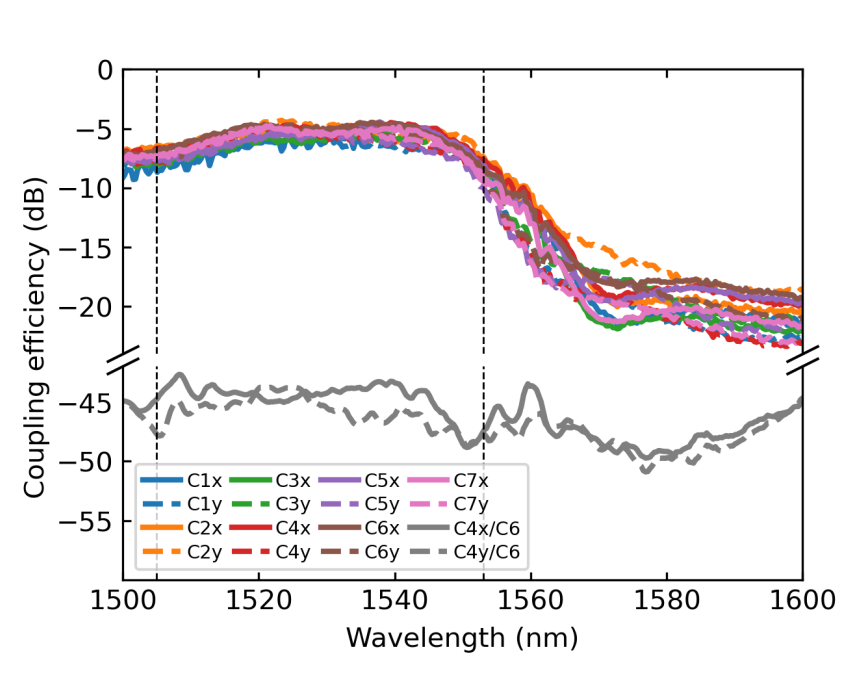}
\caption{Coupling efficiency and worst case inter-core crosstalk (between cores 4 and 6) for the dual-polarization compact couplers.
The vertical dashed lines indicate the \SI{3}{dB} bandwidth of the device.}
\label{fig:eff}
\end{figure}

The coupling efficiency measured for the dual couplers is shown in \cref{fig:eff}.
This include measurements for both polarizations in all seven cores, totaling 14 optical modes.
The plot also presents the largest measured \gls{XT} for reference, which was observed between cores 4 and 6 (see \cref{fig:2d}b).
The dual-polarization coupler shows a \SI{3}{dB} bandwidth of \SI{48}{nm}, ranging from \SI{1505}{nm} to \SI{1553}{nm}.
Compared to the numerical results presented in fig.~\ref{fig:2d} c, a shift towards shorter wavelengths can be observed. In order to elucidate this effect, complimentary simulations considering different etching heights were performed, showing that indeed, a deviation from the nominal etching height leads to a shift in the peak coupling wavelength. 
Its highest coupling efficiency is \SI{-4.3}{dB} for core 2 and the lowest efficiency is \SI{-5.4}{dB} for core 1, leading to a variation of \SI{1.1}{dB}.
This difference can be explained by the nonuniform spacing between the cores and the chip surface, as the fiber is not cleaved at an angle and no index-matching oil is used.
On the other hand, the experiments show that the maximal polarization-dependent loss is only \SI{0.54}{dB}, which can be attributed to small asymmetries in the fabricated devices and to the different bending radii of the feed waveguides.
The maximal \gls{XT} was measured at \SI{1508}{nm}: \SI{-42.7}{dB}, which is \SI{-34.7}{dB} below the coupling efficiency at this wavelength.

We have presented the design, fabrication, and characterization of compact couplers compatible with multicore fibers with a core separation distance of \SI{32}{\um}.
The coupler was designed to operate in dual-polarization, addressing the 14 modes of the fiber simultaneously.
The dual-polarization coupler has a footprint of \SI{200}{\um^2} per core, maximum coupling efficiency of \SI{-4.3}{dB}, operation bandwidth of \SI{48}{nm}, and \gls{XT} below \SI{-42.7}{dB}, representing the first experimental demonstration of dual-polarization coupling from a chip to all seven cores of a multicore fiber with an intercore distance of \SI{32}{\um}.  The most compact hexagonal arrangement of cores that could be addressed by the proposed couplers (including room for the feed waveguides) would be for an intercore distance of \SI{25}{\um}.
These figures make the device particularly attractive for advanced communication systems where wavelength division multiplexing can be combined with \gls{SDM} and polarization diversity for high data rates.
It is also worth mentioning that besides their use for multicore fibers, the designed couplers are also interesting solutions for space-constrained situations, for example in high-density fiber arrays.

\begin{backmatter}

\bmsection{Funding}
 This work was partially funded by the National Council for Scientific and Technological Development ({\smaller CNP}q) under grants 302036/2018-0, 311035/2018-3, and 432303/2018-9, and the São Paulo Research Foundation ({\smaller FAPESP}) under grants 2013/20180-3, 2015/24517-8, 2016/19270-6, and 2018/25339-4.
This study was financed in part by the Coordenação de Aperfeiçoamento de Pessoal de Nível Superior - Brasil ({\smaller CAPES}) -- Finance Code 88881.311020/2018.

\bmsection{Acknowledgments}
We thank \gls{ANT} for the support in preparation for the fabrication of the samples in a dedicated run, and we also thank Leandro Fonseca and Maicon Faria for the fruitful discussion and technical assistance.

\bmsection{Conflict of interest}
The authors declare that they have no conflicts of interest.

\bmsection{Data availability}
Data underlying the results presented in this paper are not publicly available at this time but may be obtained from the authors upon reasonable request.
\end{backmatter}

\bibliography{multicore_refs.bib}
\bibliographyfullrefs{multicore_refs.bib}
\end{document}